%% file: eps.tex
\documentclass[epj]{svjour}
\usepackage{amssymb,a4wide}
\usepackage{graphicx}
\usepackage[figuresright]{rotating}
\usepackage[small]{caption}
\renewcommand\floatpagefraction{1.}

\renewcommand\textfraction{.0}
\DeclareMathAlphabet{\mathsf}{OT1}{cmss}{m}{n}
\SetMathAlphabet{\mathsf}{bold}{OT1}{cmss}{bx}{n}

\begin{document}

\title{Indirect
Evidence for Neutralinos as Dark Matter}

\author{W. de Boer, M. Herold, C. Sander, V. Zhukov}
\institute{Institut f\"ur Experimentelle Kernphysik\\
Universit\"at Karlsruhe (TH),\\
P.O. Box 6980, 76128 Karlsruhe, Germany}

\input{config.tex}

\abstract{
From the relic density measurement by
WMAP the WIMP annihilation cross section can be determined in a
model independent way. If the WIMPS are postulated to be the
neutralinos of Supersymmetry, then only a limited region of
the supersymmetric parameter space matches this annihilation cross section.
It is shown that the resulting positrons, antiprotons and gamma rays
from the neutralino annihilation (mainly into $b\overline{b}$
quark pairs) provide the correct shape and order of magnitude for
the missing gamma and hard positron fluxes in the Galactic Models
and are consistent with antiproton production.
}
\maketitle
\section{Introduction}
Cold Dark Matter (CDM) makes up 23\% of the energy of the
universe, as deduced from the temperature anisotropies in the
Cosmic Microwave Background  in combination with data on the
Hubble expansion and the density fluctuations in the
universe~\cite{wmap}.
The nature of the CDM is unknown, but one of the most popular explanation
for it is the neutralino, a stable neutral particle predicted by
Supersymmetry~\cite{lspdm,jungman}. The neutralinos are spin 1/2
Majorana particles, which can annihilate into pairs of Standard Model (SM)
particles.
The stable decay and fragmentation products are neutrinos, photons, protons,
antiprotons, electrons and positrons. From these, the protons and
electrons are drown in the many matter particles in the universe,
but the antimatter  may be detectable above the background from nuclear
interactions, especially because of the  harder positron and gamma spectra
expected from neutralino annihilation.
This so-called indirect detection of Dark Matter has been discussed much before
(see e.g. Ref. \cite{bergstrom}). Our results differ from these previous
results by performing a statistical analysis to gamma rays,
antiprotons and gamma rays {\it simultaneously} and taking into
account the best known propagation mo\-dels and all constraints from WMAP
and electroweak data on the SUSY parameter space.
More details of this
analysis can be found in the contributed paper to this conference\cite{us}.

\begin{figure}[t] 
\begin{center}
 \includegraphics [width=0.36\textwidth,clip]{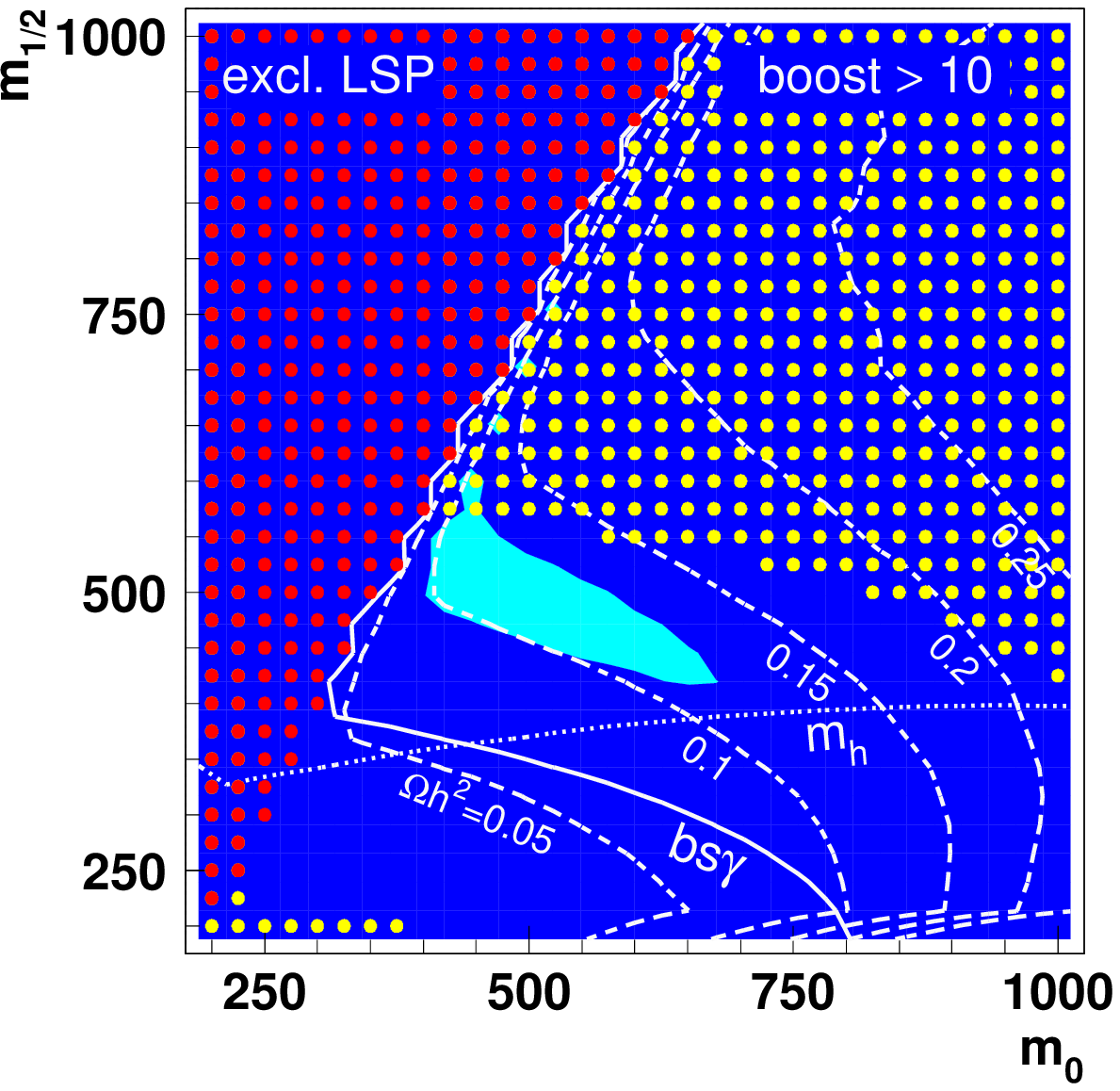}
 \includegraphics [width=0.36\textwidth,clip]{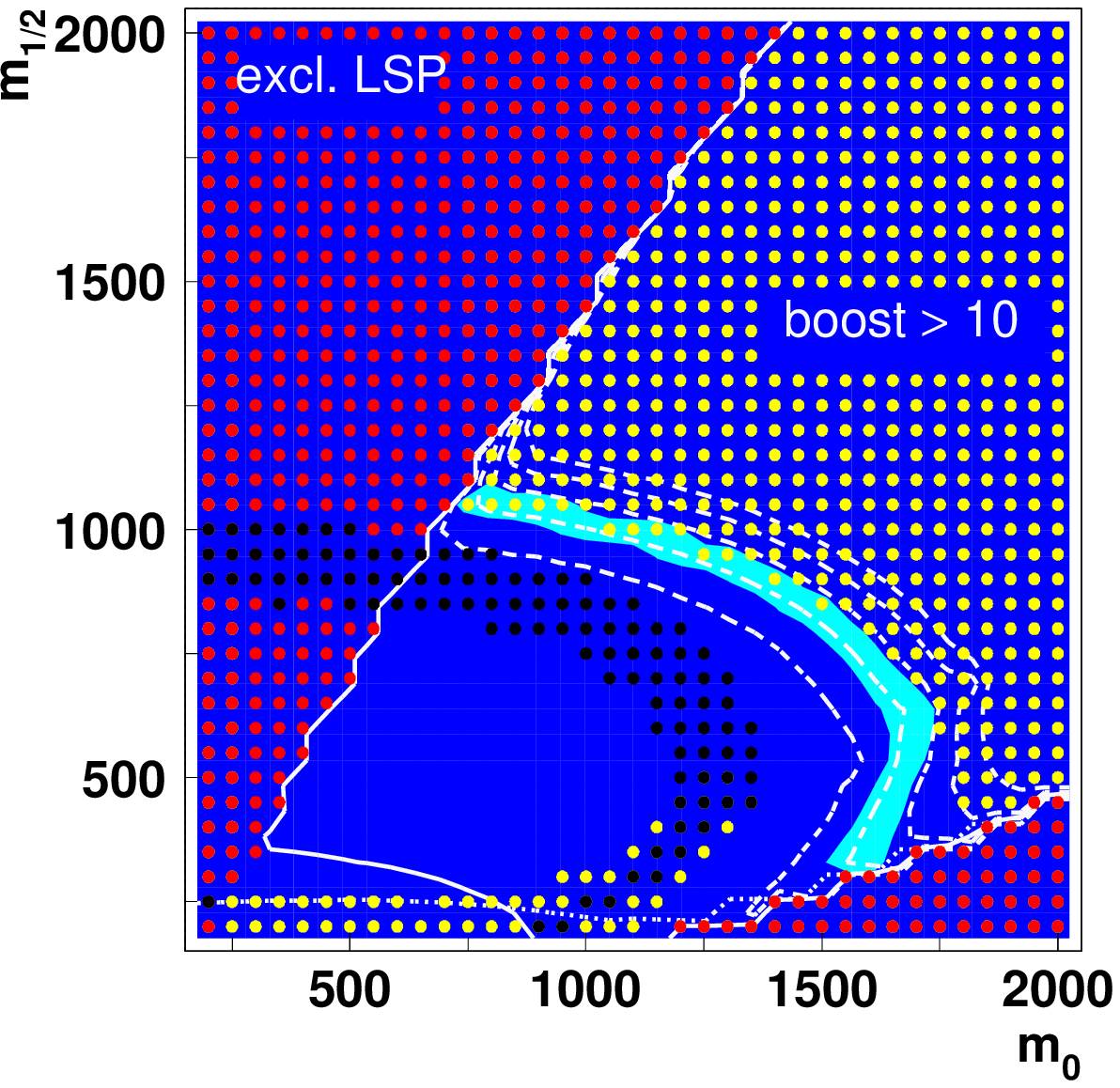}
\caption[]{\it\label{relic}
 The light shaded (blue) area is the region allowed by WMAP and the
 contours of larger $\Omega h^2$ are indicated by the dashed lines
 in steps of 0.05. The upper plot is for \tb=51 and $A_0=m_0$, while the lower
plot is for \tb=53 and $A_0=0$. For the last parameters the neutralino
annihilation hits the pseudoscalar
Higgs resonance, which allows heavier neutralinos with still a small enough
relic density.
 The black dots indicate the resonance region,
 where $|m_A-2m_{\chi_0}| \le 10$ GeV.
 The excluded regions, where the stau would be the LSP or EWSB fails or
 the boost factors are above 10 are indicated by the dots.
 }
\end{center}
\end{figure} 
\section{Annihilation Cross section Constraints from WMAP}
In the early universe all particles were produced abundantly and
were in thermal equilibrium through annihilation and production
processes.
%
At temperatures below the mass of the neutralinos the number
density drops exponentially. The annihilation rate $\Gamma=<\sigma
v> n_\chi$ drops exponentially as well, and if it drops below the
expansion rate, the neutralinos cease to annihilate
 and a relic cosmic abundance remains.
For the case that $<\sigma v>$ is energy independent, which is a
good approximation in case there is no coannihilation,
the present mass density in units of the
critical density is given by~\cite{jungman}: \bq \Omega_\chi
h^2=\frac{m_\chi n_\chi}{\rho_c}\approx (\frac{3\cdot 10^{-27}
cm^3 s^{-1}}{<\sigma v>})\label{wmap}.\eq One observes that the
present relic density is inversely proportional to the
annihilation cross section at the time of freeze out, a result
independent of the neutralino mass (except for logarithmic
corrections). For the present value of $\Omega_\chi h^2=0.11$ the
thermally averaged total cross section at the freeze-out
temperature of $m_\chi/25$ must have been $3\cdot 10^{-27}cm^3
s^{-1}$. This can be achieved only for restricted regions of
parameter space in the MSSM, as will be discussed in the next
section. Note that the annihilation cross section is given by
the Hubble expansion and therefore not dependent on the WIMP model.

\section{Predictions from Supersymmetry}
%
%
The mSUGRA model, i.e. the Minimal Supersymmetric
Standard Model (MSSM) with supergravity inspired breaking terms,
is characterized by only 5 parameters: $m_0,~m_{1/2},~\tb,~\mbox{sign}(\mu),
~A_0$\cite{susyrev}. Here $m_0$ and $m_{1/2}$ are the common masses for the
gauginos and scalars at the GUT scale, which is determined by the
unification of the gauge couplings. Exact gauge unification is still
possible with the precisely measured couplings at LEP~\cite{bs}.

The neutralinos, which are assumed to be the stable, lightest supersymmetric
particles, can annihilate through higgs- and Z-exchange in the s-channel
and SUSY particles (neutralinos, charginos, sfermions) in the t-channel.
At large values of \tb~ the dominant channel is the pseudoscalar Higgs exchange with $b\overline{b}$ quarks in the final state, which lead to a well
defined shape of the final state gammas, positrons and electrons,
since the annhihilation is practically at rest,
The regions of parameter space allowed by the WMAP data are
plotted in Fig. \ref{relic} for two values of \tb. It is
clear that for $\tb\approx 50$ only a small region is allowed.
Scanning over all possible values of \tb~
the neutralino masses allowed by the WMAP data and electroweak constraints
are in the range of 150-400 GeV\cite{us}, if we exclude the coannihilation
regions, which would lead to anomously large boost factors, as discussed
in the next section.
For the fits discussed below
we use a typical mass of 200 GeV, which corresponds to $m_{1/2}\approx 500~ GeV$.
The data are not yet sensitive enough
to distinguish between masses in the range given above.
\section{Global Fits to positrons, antiprotons and gamma rays}

Trying to disentangle the contributions from nuclear interactions
and neutralino annihilation to the antimatter fluxes and gamma
rays is in practice not easy.
We use the following strategy: the shape of the background is taken
from the GALPROP program, which represents a detailed simulation
of our galaxy\cite{galprop}. The main background of hard gammas comes
from $\pi_0$ decays, which are produced in nuclear interactions and
inverse Compton scattering of electrons on photons.
The shape of the neutralino annihilation signal is taken from
DarkSusy\cite{darksusy}. These shapes are then multiplied
by an arbitrary normalization factor, which is left as a free parameter
in the $\chi^2$ fit to the data.

\begin{figure}[t]
\begin{center}
 \includegraphics [width=0.36\textwidth,clip]{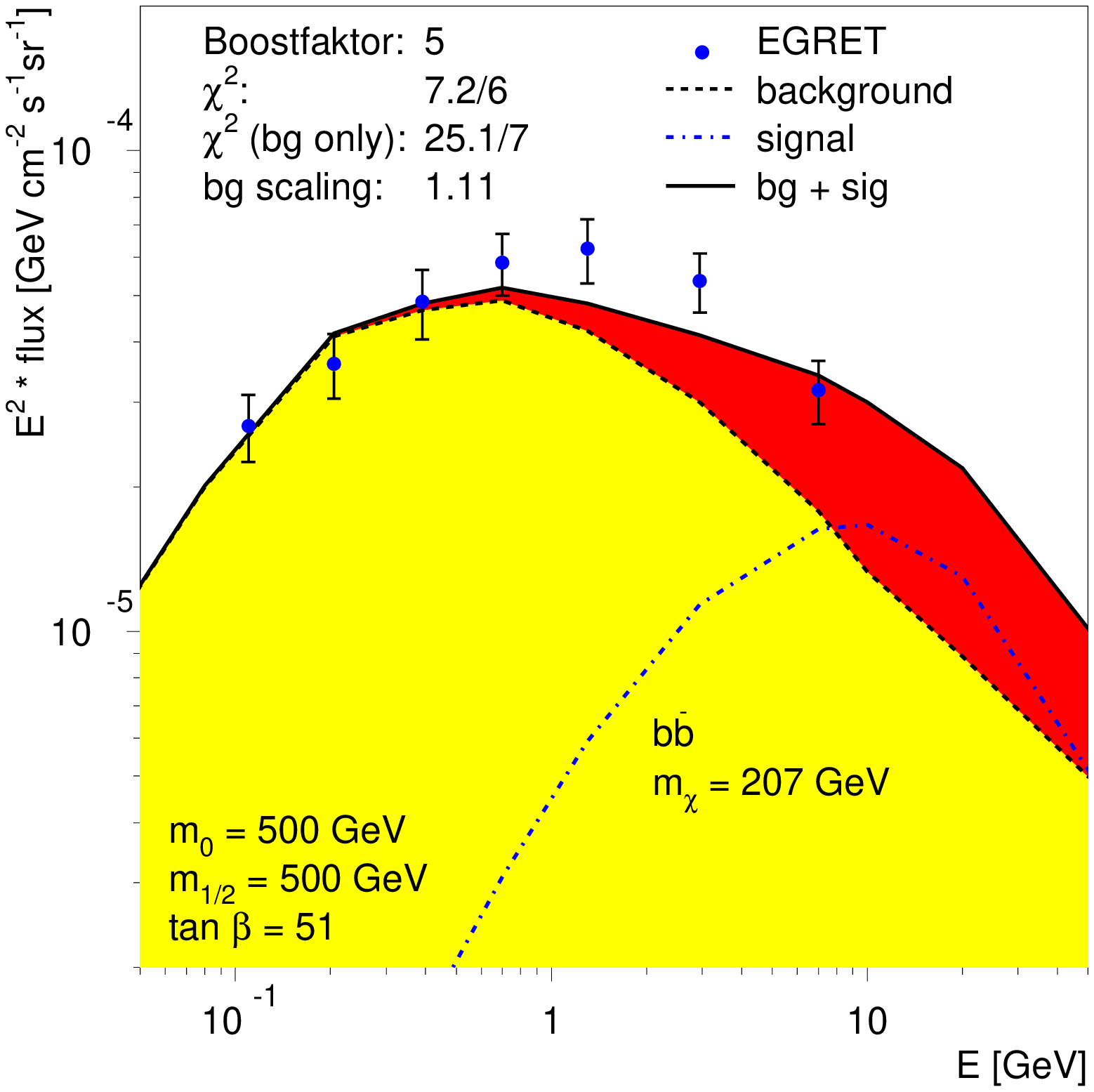}
 \includegraphics [width=0.36\textwidth,clip]{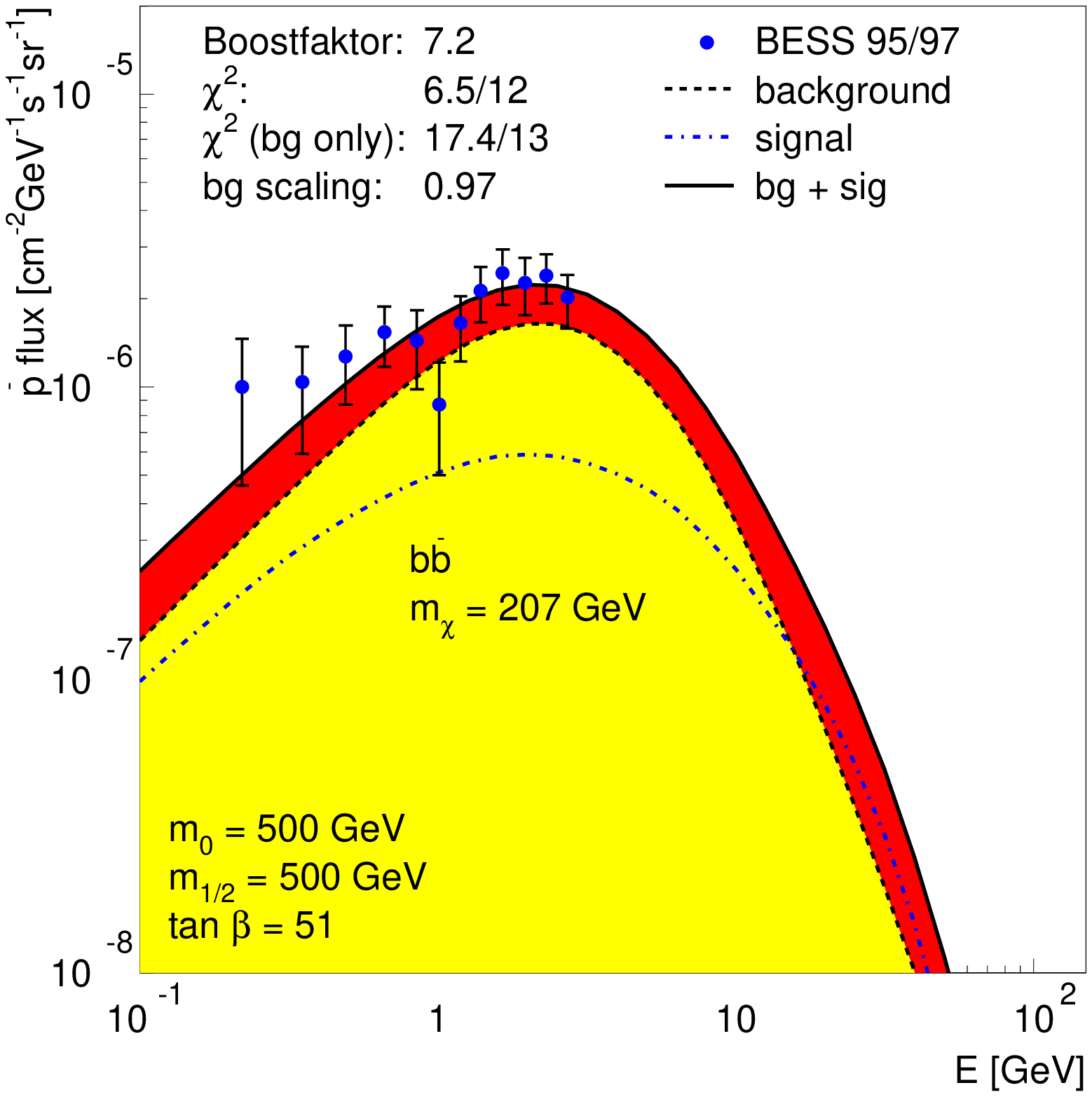}
 \includegraphics [width=0.36\textwidth,clip]{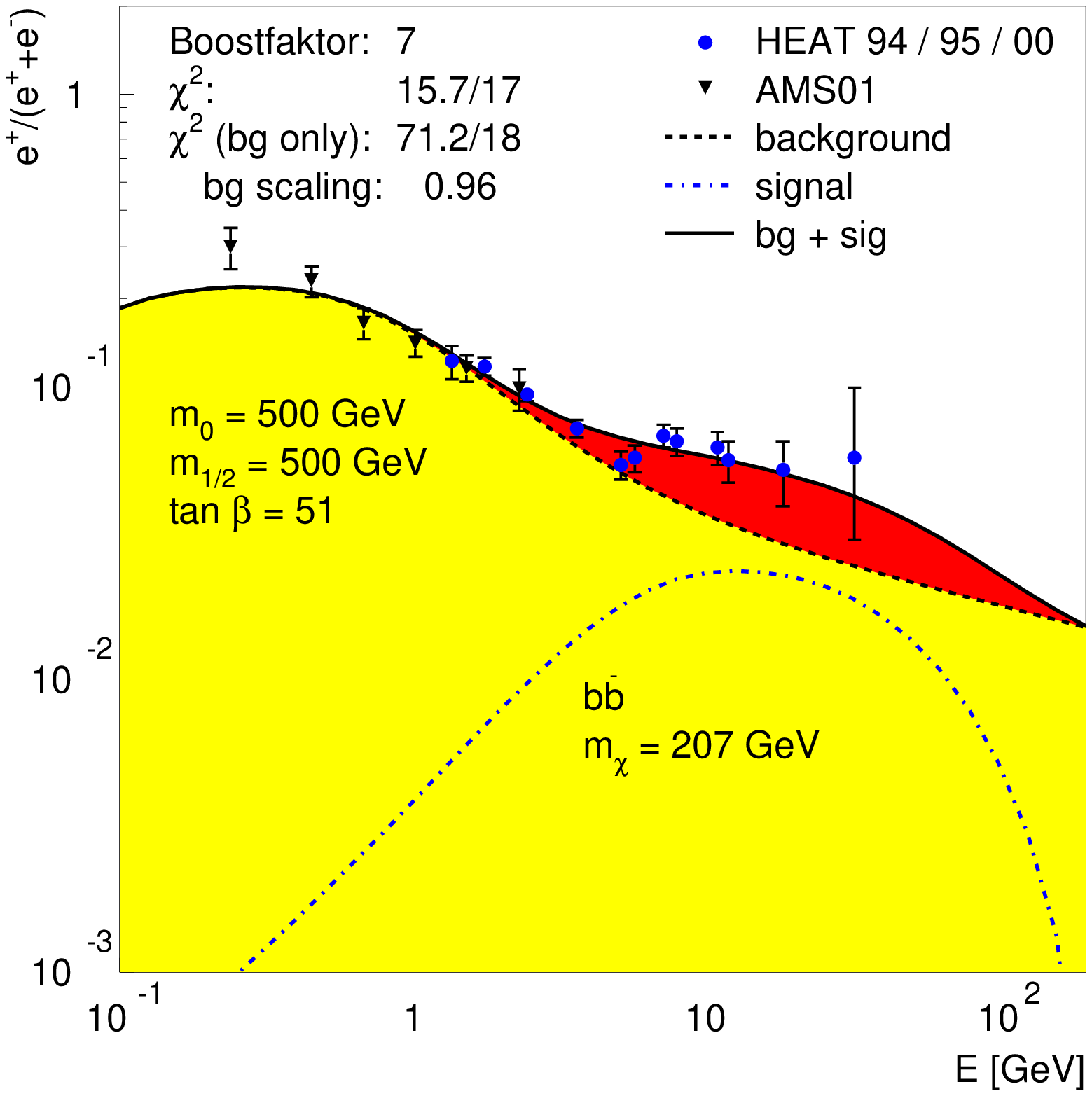}
 \caption[]{\it From top to bottom: Gamma ray , positron and antiproton
spectrum with contributions from nuclear interactions (grey/yellow)
 and neutralino annihilation (dark/red) for a neutralino mass of
 207 GeV. The normalization factors for signal, called boostfactor, and
background (bg scaling) and the values of $\chi^2)$ with and without signal
have been indicated.}
 \label{fit_gamma}
\end{center}
\end{figure}

The following data were used in the fit:
\begin{itemize}
\item
 Gamma ray data from the galactic center in the angular range
 $330^\circ<\ell<30^\circ$ and $-5^\circ<b<5^\circ$
 from the EGRET space telescope, which has
 been taking data for about 9 years on the NASA Compton Gamma Ray
 Observatory (CGRO).
 We use the data as presented in Ref.~\cite{egret}.
\item
 Positron data from AMS~\cite{ams01} and HEAT~\cite{heat}.
\item
 Antiproton data from BESS in the years 1997 and 1998~\cite{bess}
\end{itemize}

The fit results 
are shown in Fig. \ref{fit_gamma}.
The free parameters are only the normalization factors for signal 
and background for each of the particle species and their values have been
indicated in the figures.
The boost factors, i.e. the free normalization factor after correcting
for  the different propagations and energy
losses, for antiprotons, positrons and gamma rays are
all around 5-7 for the NFW halo profile\cite{nfw} taken\footnote{We use the
default $(\alpha, \beta, \gamma)$ =(1,3,1) for a scale
 $a=10$ kpc and a local relic density of 0.8 GeV/cm$^3$.}.
Much larger factors are not expected from theories of galaxy formation.
If we select SUSY parameters in the so-called coannihilation region,
where e.g. the stau and neutralino are almost degenerate, the boost
factors come out to be much larger, since the fast annihilation cross section
in the early universe by stau-neutralino coannihilation does not operate in the
present universe anymore and the small present annihilation cross section for
heavy neutralinos needs a large boost factor to fit the data.
The regions for which the boostfactors are above 10 are indicated
in Fig. \ref{relic}.

The
$\chi^2$ improves significantly with the inclusion of Dark
Matter in the fits.
The $\chi^2/d.o.f.$ is reduced from 113/35 (110/38)for the background only
fit to 29/32 (33/35) for the fit including neutralino annihilation,
where the numbers in brackets
are valid, if one takes the shape {\it and} normalization from GALPROP,
while the first
numbers are obtained if only the shape is taken and the background normalization
is left free.
This corresponds to about a 4 (6) $\sigma$ effect, if calculated with
Gaussian errors.
For the antiprotons the increase in
 probability is the  least significant, as expected, since the shape
 of background and signal are similar.

It should be noted that the statistical significance is {\it independent}
of the choice of halo or propagation parameters, since
different halo or propagation parameters would only lead to different
normalization factors in the fit, but
the $\chi^2$ is not affected, since it is only sensitive to the shape
of the distribution with free normalization parameters.

\section{Conclusion}

It is shown that the discrepancies between EGRET data
and the  galactic models can be reduced by taking as
an additional source of hard gammas the annihilation of
Dark Matter, assuming Dark Matter is made of neutralinos, as predicted
by Supersymmetry.
In addition, it is shown that adding
the positrons from neutralino annihilation in
the {\it same} Dark Matter model to the {\it same} background model
improves also the $\chi^2$ fit to the positron data significantly,
while the increase in antiprotons is compa\-tible with
the data.
These facts, 
statistical significant improvement of the global fit 
 for positrons, antiprotons and gamma
rays {\it simultaneously} for a supersymmetric model with an annihilation
cross section compatible with the model-independent WMAP value, provide
strong experimental evidence for the supersymmetric
nature of Dark Matter.

We thank V. Moskalenko and A. Strong for sha\-ring with us all
their knowledge about our galaxy, O. Reimers to provide us with
the EGRET data.
This work was supported by the DLR (Deutsches Zentrum f\"ur
Luft- und Raumfahrt).

\end{document}

%% file: config.tex

\newcommand{\unity}{\mathbf{1}}
\newcommand{\varunity}{\mbox{\rmfamily 1\hspace{-0.25em}l}}

\newcommand{\Dc}{\mathcal{D}}
\newcommand{\Hc}{\mathcal{H}}
\newcommand{\Lc}{\mathcal{L}}
\newcommand{\Oc}{\mathcal{O}}
\newcommand{\Uc}{\mathcal{U}}

\newcommand{\Rep}{\mbox{Re}}
\newcommand{\Imp}{\mbox{Im}}

\newcommand{\Rm}{\mathbb{R}}

\newcommand{\mfbox}[1]{\fbox{$\displaystyle #1$}}

\newcommand{\fnorm}[1][3/2]{\frac{1}{(2\pi)^{#1}}}

\newlength{\dslashwidth}
\newcommand{\dslash}[1]{\settowidth{\dslashwidth}{$\diagup$}\mbox{%
\hspace{0.5\dslashwidth}\makebox[0pt]{$#1$}\hspace{-0.5\dslashwidth}%
$\diagup$}}

\newcommand{\bsg}{\ensuremath{b\to X_s\gamma}}
\newcommand{\ch}{\ensuremath{\tilde{\chi}^{\pm}}}
\newcommand{\neu}{\ensuremath{\tilde{\chi}^{0}}}
\newcommand{\sinw}{\ensuremath{\sin^2\theta_W}}
\newcommand{\Mgut}{\ensuremath{M_{\mbox{\scriptsize{GUT}}}}}
\newcommand{\agut}{\ensuremath{\alpha_{\mbox{\scriptsize{GUT}}}}}
\newcommand{\tb}{\ensuremath{\tan\beta}}

\def\aii{\alpha_i^{-1}}
\def\rZ{{\rm Z}}
\def\rW{{\rm W}}
\def\rG{{\rm GUT}}
\def\rS{{\rm SUSY}}
\def\rH{{\rm Higgs}}
\def\rF{{\rm Fam}}
\def\MG{M_\rG}
\renewcommand{\floatpagefraction}{0.05}
\renewcommand{\textfraction}{0.05}
\newcommand{\mc}{Monte Carlo }
\newcommand{\mcs}{Monte Carlos }
\newcommand{\brem}{brems\-strah\-lung }
\newcommand{\bq}{\begin{equation}}
\newcommand{\eq}{\end{equation}}
\newcommand{\ba}{\begin{array}}
\newcommand{\ea}{\end{array}}
\newcommand{\bqa}{\begin{eqnarray}}
\newcommand{\eqa}{\end{eqnarray}}
\newcommand{\nn}{\nonumber \\}
\newcommand{\mpmm}{\mu^{+}\mu^{-}}
\newcommand{\tptm}{\tau^{+}\tau^{-}}
\newcommand{\sq}{^{2}}
\newcommand{\etal}{\it et al.\rm}
\newcommand{\ra}{\rightarrow}
\newcommand{\lnf}{{\ifmmode \Lambda^{(N_f)} \else $\Lambda^{(N_f)}$\fi}}
\newcommand{\ms}{{\ifmmode \overline{MS} \else $\overline{MS}$\fi}}
\newcommand{\dr}{{\ifmmode \overline{DR} \else $\overline{DR}$\fi}}
\newcommand{\lms}{{\ifmmode \Lambda^{(5)}_{\overline{MS}} \else $\Lambda^{(5)}_{\overline{MS}}$\fi}}
\newcommand{\lam}{{\ifmmode \Lambda \else $\Lambda$\fi}}
\newcommand{\gev}{{\ifmmode {\rm GeV} \else ${\rm GeV}$\fi}}
\newcommand{\gevc}{{\ifmmode {\rm GeV/c^2} \else ${\rm GeV/c^2}$\fi}}
\newcommand{\tev}{{\ifmmode {\rm TeV} \else ${\rm TeV}$\fi}}
\newcommand{\tevc}{{\ifmmode {\rm TeV/c^2} \else ${\rm TeV/c^2}$\fi}}
\newcommand{\lp}{{\ifmmode L^+  \else $L^+$\fi}}
\newcommand{\lm}{{\ifmmode L^-  \else $L^-$\fi}}
\newcommand{\mlp}{{\ifmmode M(L^-) \else $M(L^-)$\fi}}
\newcommand{\mlz}{{\ifmmode M(L^0) \else $M(L^0)$\fi}}
\newcommand{\lz}{{\ifmmode L^0 \else $L^0$\fi}}
\newcommand{\ev}{{\ifmmode GeV/c^2 \else $GeV/c^2$\fi}}
\newcommand{\tri}{{\ifmmode \triangleup \else $\triangleup$\fi}}
\newcommand{\unl}{{\ifmmode U_{lL^0} \else $U_{lL^0}$\fi}}\newcommand{\gL}{{\ifmmode g_L \else $g_{L}$\fi}}
\newcommand{\gR}{{\ifmmode g_R  \else $g_{R}$\fi}}
\newcommand{\gumu}{{\ifmmode \gamma^{\mu} \else $\gamma^{\mu}$\fi}}
\newcommand{\gunu}{{\ifmmode \gamma^{\nu} \else $\gamma^{\nu}$\fi}}
\newcommand{\gdmu}{{\ifmmode \gamma_{\mu} \else $\gamma_{\mu}$\fi}}
\newcommand{\gdnu}{{\ifmmode \gamma_{\nu} \else $\gamma_{\nu}$\fi}}
\newcommand{\stw}{{\ifmmode\sin^2\theta_W \else $\sin^{2}\theta_{W}$ \fi}}
\newcommand{\sws}{{\ifmmode \;\sin^2\theta_W  \else $\;\sin^{2}\theta_{W}$ \fi}}
\newcommand{\cws}{{\ifmmode \;\cos^2\theta_W  \else $\;\cos^{2}\theta_{W}$ \fi}}
\newcommand{\sw}{{\ifmmode \;\sin\theta_W  \else $\sin\theta_{W}$ \fi}}
\newcommand{\cw}{{\ifmmode \;\cos\theta_W  \else $\;\cos\theta_{W}$ \fi}}
\newcommand{\tw}{{\ifmmode \;\tan\theta_W  \else $\;\tan\theta_{W}$ \fi}}
\newcommand{\qq}{{\ifmmode q\overline{q} \else $q\overline{q}$\fi}}
\newcommand{\lR}{{\ifmmode l_R \else $l_R$\fi}}
\newcommand{\lL}{{\ifmmode l_L \else $l_L$\fi}}
\newcommand{\nt}{{\ifmmode \nu_{\tau} \else $\nu_{\tau}$\fi}}
\newcommand{\nuR}{{\ifmmode \nu_R  \else $\nu_R$\fi}}
\newcommand{\nuL}{{\ifmmode \nu_L  \else $\nu_L$\fi}}
\newcommand{\qR}{{\ifmmode g_R  \else $q_R$\fi}}
\newcommand{\qL}{{\ifmmode q_L  \else $q_L$\fi}}
\newcommand{\qRp}{{\ifmmode q_R'  \else $q_{R}$'\fi}}
\newcommand{\qLp}{{\ifmmode q_L'  \else $q_{L}$'\fi}}
\newcommand{\est}{{\ifmmode e^{\bf \ast} \else $e^{\bf \ast}$\fi}}
\newcommand{\lst}{{\ifmmode l^{\bf \ast} \else $l^{\bf \ast}$\fi}}
\newcommand{\must}{{\ifmmode \mu^{\bf \ast} \else $\mu^{\bf \ast}$\fi}}
\newcommand{\taust}{{\ifmmode \tau^{\bf \ast} \else $\tau^{\bf \ast}$ \fi}}
\newcommand{\pperp}{{\ifmmode p_t  \else $p_t$\fi}}
\newcommand{\et}{{\ifmmode E_t  \else $E_t$\fi}}
\newcommand{\xt}{{\ifmmode x_t  \else $x_t$\fi}}
\newcommand{\smumu}{{\ifmmode \sigma_{\mu\mu}  \else $\sigma_{\mu\mu}$ \fi}}
\newcommand{\eg}{{\ifmmode e\gamma  \else $e\gamma$\fi}}
\newcommand{\epem}{{\ifmmode e^+e^-  \else $e^+e^-$\fi}}
\newcommand{\lplm}{{\ifmmode L^+L^-  \else $L^+L^-$\fi}}
\newcommand{\pp}{{\ifmmode p\overline p  \else $p\overline p$\fi}}
\newcommand{\llz}{{\ifmmode L^0\overline{L}^0 \else $L^0\overline{L}^0$\fi}}
\newcommand{\epemt}{{\ifmmode e^+e^- \to  \else $e^+e^- \to$\fi}}
\newcommand{\eb}{{\ifmmode E_{beam}  \else $E_{beam}$\fi}}
\newcommand{\ip}{{\ifmmode pb^{-1}  \else $pb^{-1}$\fi}}
\newcommand{\upm}{{\ifmmode ^{\pm}  \else $^{\pm}$\fi}}
\newcommand{\de}{{\ifmmode ^{\circ}  \else $^{\circ}$ \fi}}
\newcommand{\appr}{{\ifmmode \sim \else $\sim$ \fi}}
\newcommand{\corresp}{{\ifmmode \stackrel{\wedge}{=} \else $\stackrel{\wedge}{=}$ \fi}}
\newcommand{\sqrts}{{\ifmmode \sqrt{s} \else $\sqrt{s}$\fi}}
\newcommand{\zz}{{\ifmmode Z^0  \else $Z^0$\fi}}
\newcommand{\mz}{{\ifmmode M_{Z}  \else $M_{Z}$\fi}}
\newcommand{\mzs}{{\ifmmode M_{Z}^2  \else $M_{Z}^2$\fi}}
\newcommand{\mw}{{\ifmmode M_{W}  \else $M_{W}$\fi}}
\newcommand{\mws}{{\ifmmode M_{W}^2  \else $M_{W}^2$\fi}}
\newcommand{\mh}{{\ifmmode M_{Higgs}  \else $M_{Higgs}$\fi}}
\newcommand{\gt}{{\ifmmode \Gamma_{tot} \else $\Gamma_{tot}$\fi}}
\newcommand{\msusy}{{\ifmmode M_{SUSY}  \else $M_{SUSY}$\fi}}
\newcommand{\msusys}{{\ifmmode M_{SUSY}^2  \else $M_{SUSY}^2$\fi}}
\newcommand{\su}{{\ifmmode SU(3)_C\otimes\- SU(2)_L\otimes\- U(1)_Y \else $SU(3)_C\otimes SU(2)_L\otimes U(1)_Y$\fi}}
\newcommand{\suthree}{{\ifmmode SU(3)_C  \else $SU(3)_C$\fi}}
\newcommand{\sutwo}{{\ifmmode  SU(2)_L\otimes U(1)_Y \else $SU(2)_L\otimes U(1)_Y$\fi}}
\newcommand{\taup} {{\ifmmode \tau_{proton} \else $\tau_{proton}$\fi}}
\newcommand{\as}{{\ifmmode \alpha_{s}  \else $\alpha_{s}$\fi}}
\newcommand{\mgut}{{\ifmmode M_{GUT}  \else $M_{GUT}$\fi}}
\newcommand{\mguts}{{\ifmmode M_{GUT}^2  \else $M_{GUT}^2$\fi}}
\newcommand{\mze} {{\ifmmode m_0        \else $m_0$\fi}}
\newcommand{\mha}{{\ifmmode m_{1/2}    \else $m_{1/2}$\fi}}
\newcommand{\mb} {{\ifmmode m_{b}    \else $m_{b}$\fi}}
\newcommand{\mt} {{\ifmmode m_{t}    \else $m_{t}$\fi}}
\newcommand{\mts} {{\ifmmode m_{t}^2    \else $m_{t}^2$\fi}}
\newcommand {\rb}[1]{\raisebox{1.5ex}[-1.5ex]{#1}}
\newcommand{\mtau}{{\ifmmode m_{\tau}  \else $m_{\tau}$\fi}}
\newcommand{\dpp}{{\ifmmode \delta_{pert} \else $\delta_{pert}$\fi}}
\newcommand{\dnp}{{\ifmmode\delta_{non-pert}\else$\delta_{non-pert}$\fi}}
\newcommand{\dew}{{\ifmmode \delta_{\rm EW}\else $\delta_{\rm EW}$\fi}}
\newcommand{\rt}{{\ifmmode R_{\tau}  \else $R_{\tau} $\fi}}
\newcommand{\rz}{{\ifmmode R_{Z}  \else $R_{Z} $\fi}}
\newcommand{\into}{\rightarrow}
\newcommand{\SM}{Standard Model}
\newcommand{\swb}{{\ifmmode \sin^2\theta_{\overline{MS}} \else $\sin^2\theta_{\overline{MS}}$\fi}}
\newcommand{\cwb}{{\ifmmode \cos^2\theta_{\overline{MS}} \else $\cos^2\theta_{\overline{MS}}$\fi}}
\newcommand{\ttbs}{\char'134}
\newcommand{\AmS}{{\protect\the\textfont2 A\kern-.1667em\lower.5ex\hbox{M}\kern-.125emS}}
\newcommand{\besg}{$b  \to  X_s \gamma~ $}
\newcommand{\mzero}{\rm m_0}
\newcommand{\mhalf}{\rm m_{1/2}}